\documentclass[pre,twocolumn,aps,eqsecnum]{revtex4}
\usepackage{epsfig}

\begin{document}

\title{Yielding Transitions and Grain-Size Effects in Dislocation Theory} 
\author{J.S. Langer}
\affiliation{Department of Physics, University of California, Santa Barbara, CA  93106-9530}

\date{\today}

\begin{abstract}
The statistical-thermodynamic dislocation theory developed in previous papers is used here in an analysis of yielding transitions and grain-size effects in polycrystalline solids.  Calculations are based on the 1995 experimental results of Meyers et al. for polycrystalline copper under strain-hardening conditions.  The main assertion is that the well known Hall-Petch effects are caused by enhanced strengths of dislocation sources at the edges of grains instead of the commonly assumed resistance to dislocation flow across grain boundaries.  The theory describes rapid transitions between elastic and plastic deformation at yield points;  thus it can be used to predict grain-size dependence of both yield stresses and flow stresses.  
\end{abstract}

\maketitle

\section{Introduction}
\label{Intro} 

Here is another chapter in my recent efforts to develop a theory of polycrystalline plasticity based on the principles of nonequilibrium statistical thermodynamics.  There is much that remains to be done to complete this project; but I hope to make it clear that the next steps need to be experimental.

The preceding papers in this series, \cite{LBL-10,JSL-15,JSL-16,JSL-16a} , are based on two unconventional ideas.  The first of these is that, under nonequilibrium conditions, the atomically slow configurational degrees of freedom of deforming solids are characterized by an effective disorder temperature that is substantially different from the ordinary thermal temperature.  These two temperatures play analogous roles in the sense that both are thermodynamically well defined dynamical variables whose equations of motion determine the irreversible behaviors of these systems.  The second  principal idea is that entanglement of dislocations is the overwhelmingly dominant cause of resistance to deformation in polycrystalline materials.  These two ideas have led to successfully predictive theories of strain hardening, steady-state stresses over exceedingly wide ranges of strain rates, and adiabatic shear banding.  
 
Two related subjects that so far have been touched on only briefly in the preceding papers are the nature of yielding transitions and the roles played by the grain size. I use the term ``yield stress'' here in the conventional way.  That is, ``yield stress'' denotes the minimum stress required to cause a material to deform plastically, implying that the material deforms only elastically below that stress.  However, this  term is often used more generally in the literature to denote the flow stress. I will not do that here because I want to focus on yielding as an important dynamical phenomenon that needs to be understood by itself.

The Hall-Petch formula  \cite{HALL-51,PETCH-53} describing the effects of grain size was first  published in 1951.  It generally is written:
\begin{equation}
\label{HPeqn}
\sigma = \sigma_0 + {k_s\over \sqrt{d}},
\end{equation}
where $\sigma$ is a measured stress, $d$ is the average grain diameter, and $\sigma_0$ and $k_s$ are fitting parameters.  In the 2014 review by Armstrong  \cite{ARMSTRONG-14}, and in the 1995 paper by Meyers et al. \cite{MEYERSetal-95} on which all of the following analysis is based, $\sigma_0$ is said to be a ``frictional'' stress.  By using this term, these authors imply that $\sigma_0$ is a generalization of the Peierls-Nabarro drag stress that resists the motion of dislocations when they are moving freely through a crystal.  In other recent reviews such as those by Armstrong et al. \cite{ARMSTRONGetal-09} and Gray \cite{GRAY-12}, $\sigma_0$ becomes a function of strain rate and temperature that purportedly describes  stress-strain relations more generally.  I have learned a great deal from the 2003 review by Kocks and Mecking. \cite{KOCKS-MECKING-03} These authors agree with Cottrell's famous assertion  \cite{COTTRELL-02} that a true theory of strain hardening is beyond the range of theoretical physics; but they speak optimistically about phenomenological models as the best possible alternatives.  I disagree, and  will continue to argue here that a physics-based approach is feasible and absolutely essential.   

Hall and Petch, and almost everyone else working in this field for the last sixty years, have interpreted the term proportional to $d^{-1/2}$ on the right-hand side of Eq.(\ref{HPeqn}) to mean that dislocation motion is impeded at grain boundaries, i.e. that the dislocations ``pile up'' (see  \cite{ARMSTRONG-14}) at those places and measureably increase the stress required to move dislocations across the system as a whole.  In 1946, Zener  \cite{ZENER-46} may have been the first to point out that stress concentration factors proportional to $d^{-1/2}$, near shear cracks or other obstacles with length scales $d$, may be relevant to dislocation dynamics. Thus, it has long seemed reasonable to suppose that the HP formula describes a combination of drag and grain-boundary forces opposing dislocation motion. 

It is unclear to me, however, why these two resistive mechanisms should appear additively in the HP equation if, indeed, that equation is fundamentally a relation between stress and plastic flow.  In general, flows are governed by what Cottrell called ``the weakest links''  \cite{COTTRELL-02} or, equivalently, the narrowest bottlenecks.  According to  \cite{LBL-10,JSL-15,JSL-16,JSL-16a}, by far the narrowest of these bottlenecks are the thermally activated processes by which entangled dislocations are depinned from one another.  For example, in \cite{LBL-10} we showed that the times taken for depinned dislocations to move to their next pinning sites are completely negligible in comparison with the pinning times, so that the Peierls-Nabarro drag forces disappear from relations between stress and strain rate in most experimentally interesting situations.   

What, then, is the physical meaning of the HP formula?  I think that the answer to this question comes directly and unambiguously from the experimental data of Meyers et al  \cite{MEYERSetal-95}.  In the thermodynamic equation of motion for the dislocation density, Eq.(\ref{rhodot}) below, the factor $\kappa_1$ is proportional to the fraction of the input power that is converted into the formation energy of new dislocations.  As I will show, these experiments tell us  that $\kappa_1$ -- and only $\kappa_1$ to any substantial degree -- has the HP form shown in Eq.(\ref{HPeqn}).  There may be many different kinds of dislocation sources in these systems; but, to a first approximation, it seems safe to assume that these sources are independent of each other and, therefore, appear additively in the formula for $\kappa_1$. Apparently, a substantial number of these sources occur on grain boundaries.  The main theme of this paper is that this term, by itself, accounts for the Hall-Petch-like behaviors. 

In what follows, I start by briefly restating the equations of motion for the relevant dynamic state variables as given in \cite{JSL-16a}.  I then revisit my earlier analysis  \cite{JSL-15} of the data from \cite{MEYERSetal-95}, largely to exhibit the $d$ dependence of $\kappa_1$. Finally, by using this theory to simulate a variety of loading histories, I demonstrate how HP behaviors appear in measurements of yield stresses and flow stresses. 

\section{Equations of Motion}
\label{EOM}

Consider a strip of polycrystalline material, of width $2\,W$, oriented in the $x$ direction, being driven in simple shear at velocities $V_x$ and $-V_x$ at its top and bottom edges. Let $y$ denote the transverse position. perpendicular to the $x$ axis.  The total strain rate is $V_x/W \equiv Q/\tau_0$, where $\tau_0 = 10^{-12} s$ is a characteristic microscopic time scale. The local, elastic plus plastic strain rate is $\dot\epsilon(y) = dv_x/dy$, where $v_x$ is the material velocity in the $x$ direction. This motion is driven by a time dependent, spatially uniform, shear stress $\sigma$. Because this system is undergoing steady-state shear, we can replace the time $t$ by the accumulated total strain, say $\epsilon$, so that $\tau_0\,\partial/\partial t \to Q\,\partial/\partial \epsilon$. Then denote the dimensionless, possibly $y$-dependent, plastic strain rate by $q(y,\epsilon) \equiv \tau_0\,\dot\epsilon^{pl}(y,\epsilon)$.

The internal state variables that describe this system are the areal density of dislocations $\rho\equiv  \tilde\rho/ b^2$ (where $b$ is the length of the Burgers vector), the effective temperature $\tilde\chi$ (in units of a characteristic dislocation energy $e_D$), and the ordinary temperature $\tilde\theta$ (in units of the pinning temperature $T_P = e_P/k_B$, where $e_P$ is the pinning energy defined below).  Note that  $1/\sqrt{\rho}$ is the average distance between dislocations. All three of these dimensionless quantities, $\tilde\rho$, $\tilde\chi$, and $\tilde\theta$, are functions of $\epsilon$ and $y$. 

The central, dislocation-specific ingredient of this analysis is the thermally activated depinning formula for the dimensionless plastic strain rate $q$ as a function of a non-negative stress $\sigma$:  
\begin{equation}
\label{qdef}
q(y,\epsilon) = \sqrt{\tilde\rho} \,\exp\,\Bigl[-\,{1\over \tilde\theta}\,e^{-\sigma/\sigma_T(\tilde\rho)}\Bigr]. 
\end{equation}
Here, $\sigma_T(\tilde\rho)= \mu_T\,\sqrt{\tilde\rho}$ is the Taylor stress, and $\mu_T \cong \mu/31$, where $\mu$ is  the elastic shear modulus.  The pinning energy $e_P$ is large, of the order of electron volts, so that $\tilde\theta$ is very small.  As a result, $q(y,\epsilon)$ is an extremely rapidly varying function of $\sigma$ and $\tilde\theta$.  This strongly nonlinear behavior is the key to understanding yielding transitions as well as many other important features of polycystalline plasticity.  For example, the extremely slow variation of the steady-state stress as a function of strain rate discussed in  \cite{LBL-10} is the converse of the extremely rapid variation in Eq.(\ref{qdef}). 

The equation of motion for the scaled dislocation density $\tilde\rho$ describes energy flow. It says that some fraction of the power delivered to the system by external driving is converted into the energy of dislocations, and that that energy is dissipated according to a detailed-balance analysis involving the effective temperature $\tilde\chi$.  This equation is: 
\begin{equation}
\label{rhodot}
{\partial\tilde\rho\over \partial\epsilon} = \kappa_1\,{\sigma\,q\over \nu_0^2\,\mu_T\,Q}\, \Bigl[1 - {\tilde\rho\over \tilde\rho_{ss}(\tilde\chi)}\Bigr],
\end{equation}
where $\tilde\rho_{ss}(\tilde\chi) = e^{- 1/\tilde\chi}$ is the steady-state value of $\tilde\rho$ at given $\tilde\chi$.  As stated earlier, much of the physics of this equation is contained in the coefficient $\kappa_1$, which is proportional to an energy conversion factor, and is given in terms of the hardening rate at the onset of plastic flow by the relation
\begin{equation}
\label{Theta0}
\kappa_1 = {2\over \mu_T} \left({\partial \sigma\over \partial\epsilon}\right)_{onset}.
\end{equation} 
The other quantity that appears in the prefactor in Eq.(\ref{rhodot}) is
\begin{equation}
\label{nudef}
\nu_0\equiv \ln\Bigl({1\over \tilde\theta}\Bigr) - \ln\Bigl[\ln\Bigl({\sqrt{\tilde\rho_0}\over Q}\Bigr)\Bigr],
\end{equation}
where $\tilde\rho_0$ is the value of $\tilde\rho$ at onset.  Because this quantity appears here only as the argument of a double logarithm, it is best approximated for computational purposes just by $\tilde\rho$ itself. See \cite{JSL-16a} for a more detailed derivation of Eq.(\ref{rhodot}). 

\begin{figure}[here]
\centering \epsfig{width=.5\textwidth,file=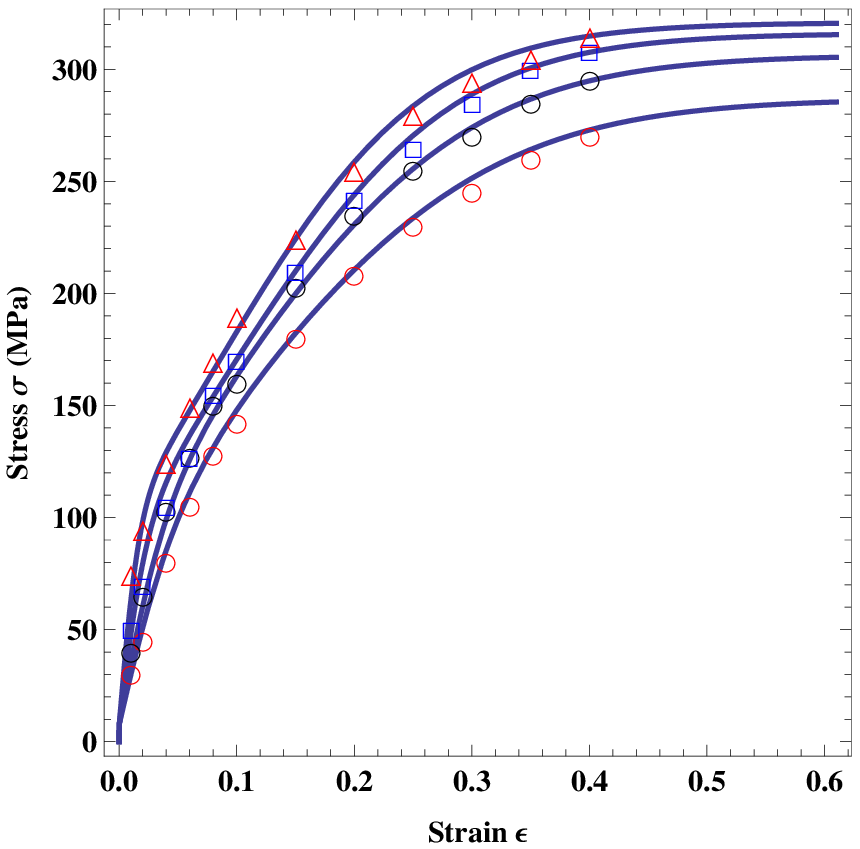} \caption{(Color online) Theoretical stress-strain curves for polycrystalline copper at the small strain rate $\dot\epsilon = 10^{-3}\,s^{-1}$, for grain diameters $d = 9.5,\,25,\,117$ and $315~m\mu$, shown from top to bottom.  The experimental points are taken from \cite{MEYERSetal-95}.} \label{HPFig1}
\end{figure}

The equation of motion for the scaled effective temperature $\tilde\chi$ is a statement of the first law of thermodynamics for the configurational subsystem: 
\begin{equation}
\label{chidot}
{\partial\,\tilde\chi\over \partial\epsilon} = \kappa_2\,{\sigma\,q\over \mu_T\,Q}\,\Bigl( 1 - {\tilde\chi\over \tilde\chi_0} \Bigr). 
\end{equation}
Here, $\tilde\chi_0$ is the steady-state value of $\tilde\chi$ for strain rates appreciably smaller than inverse atomic relaxation times, i.e. much smaller than $\tau_0^{-1}$. The overall, dimensionless factor $\kappa_2$ is inversely proportional to the effective specific heat $c_{e\!f\!f}$. Unlike $\kappa_1$, whose value can be determined directly from experiment via Eq.(\ref{Theta0}), $\kappa_2$ must be determined on a case by case basis by fitting the data. I have omitted a term on the right-hand side of Eq.(\ref{chidot}) that accounts for storage of energy in the form of dislocations.  In \cite{JSL-15}, I thought that this term might be significant, but I now think that it is not relevant for present purposes.

The equation of motion for the scaled, ordinary temperature $\tilde\theta$ is the usual thermal diffusion equation with a source term proportional to the input power.  I assume that, of the three state variables, only $\tilde\theta$ may diffuse in the spatial dimension $y$. Thus,
\begin{equation}
\label{thetadot}
{\partial\tilde\theta\over \partial\epsilon} = K\,{\sigma\,q\over Q} + {K_1\over Q}\,{\partial^2 \tilde\theta\over \partial\,y^2} - {K_2\over Q}\,(\tilde\theta - \tilde\theta_0).
\end{equation} 
Here, $K = \beta/ (T_P\,c_p\,\rho_d)$, where $c_p$ is the thermal heat capacity per unit mass, $\rho_d$ is the mass density, and $0< \beta < 1$ is a dimensionless conversion factor. $K_1$ is proportional to the thermal diffusion constant, and $K_2$ is a thermal transport coefficient that assures that the system  remains close to the ambient temperature $\tilde\theta_0 = T_0/T_P$ under slow deformation, i.e. small $Q$.

\begin{figure}[here]
\centering \epsfig{width=.5\textwidth,file=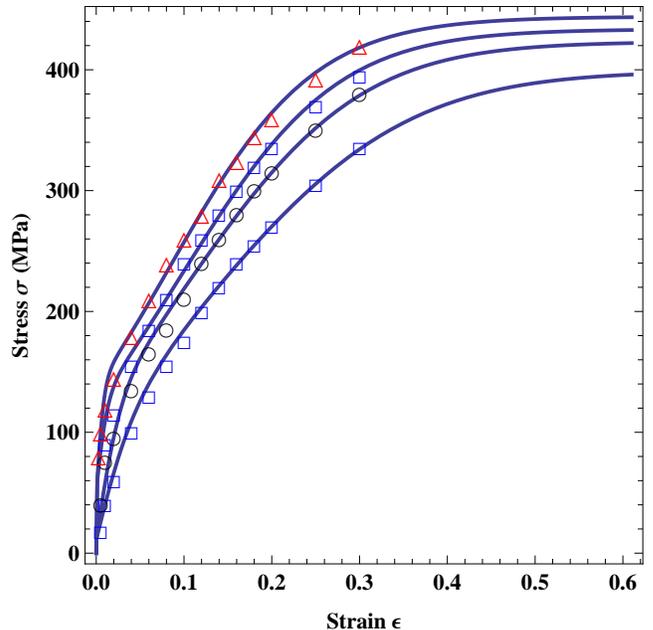} \caption{(Color online) Theoretical stress-strain curves for polycrystalline copper at the large strain rate $\dot\epsilon = 3 \times 10^3\,s^{-1}$, for grain diameters $d = 9.5,\,25,\,117$ and $315~m\mu$, shown from top to bottom.  The experimental points are taken from \cite{MEYERSetal-95}. } \label{HPFig2}
\end{figure}

It remains to write an equation of motion for the stress $\sigma(\epsilon)$ which, to a very good approximation, should be independent of position $y$ for this model of simple shear.  I start, therefore, with the  local relation  $\dot\sigma = \mu[\dot\epsilon(y) - \dot\epsilon^{pl}(y)]$, which  becomes 
\begin{equation}
\label{sigmaeqn}
{d\sigma\over d \epsilon} = \mu\,\left[{\tau_0\over Q}\,{dv_x\over dy} - {q(y,\epsilon)\over Q}\right].
\end{equation}
One simple strategy for enforcing spatial uniformity of $\sigma$ is to integrate both sides of this relation over $y$ and divide by $2 W$ to find
\begin{equation}
\label{sigmaeqn1}
{d\sigma\over d \epsilon} = \mu\,\left[1 -\int_{-W}^{+W}\,{dy\over 2W}\,{q(y,\epsilon)\over Q}\right].
\end{equation}
An even simpler strategy for numerical purposes is to replace Eq.(\ref{sigmaeqn1}) by
\begin{equation}
\label{sigmaeqn2}
{\partial\sigma\over \partial\epsilon} = \mu\,\left[1-{q(y,\epsilon)\over Q}\right] + M\,{\partial^2 \sigma\over \partial y^2},
\end{equation}
and to use a large enough value of the ``diffusion constant'' $M$ that $\sigma$ remains constant as a function of $y$. I have chosen $M = 10^5$, and have checked by direct comparisons with the predictions of Eq.(\ref{sigmaeqn1}) that this procedure is accurate. I also have set $W = 1$ in order to define the length scale. 

\begin{figure}[here]
\centering \epsfig{width=.5\textwidth,file=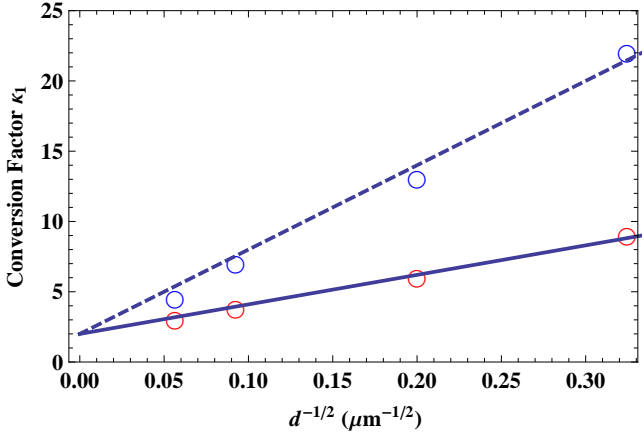} \caption{(Color online) Hall-Petch plots for the conversion factor $\kappa_1$ as a function of grain diameter $d$.  The lower solid curve is for the small strain rate, $\dot\epsilon = 10^{-3}\,s^{-1}$, and the upper dashed curve is for $\dot\epsilon = 3 \times 10^3\,s^{-1}$.  These curves are fit by the formulas shown in Eq.(\ref{HPkappa1}) . } \label{HPkappa}
\end{figure}

\begin{figure}[here]
\centering \epsfig{width=.5\textwidth,file=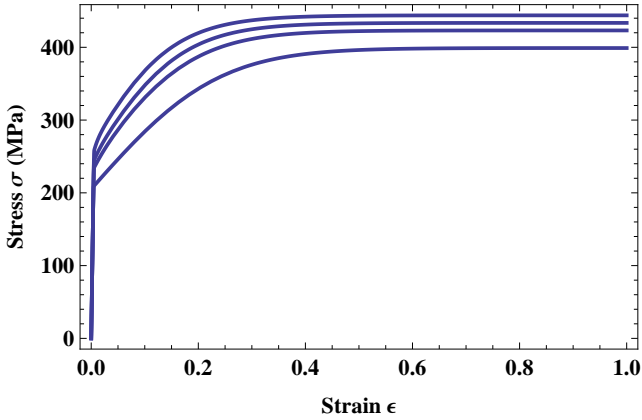} \caption{Stress-strain curves for prehardened isothermal samples. The grain diameters are $d = 9.5,\,25,\,117$ and $315~m\mu$ from top to bottom  } \label{HPyield1}
\end{figure}

\section{Grain-Size Dependence of the Hardening Curves}

In \cite{JSL-15}, I showed how the preceding equations of motion can be used to analyze the data of Meyers et al. \cite{MEYERSetal-95}, who report measurements of stress-strain curves for polycrystalline copper at two very different strain rates, $\dot\epsilon = 10^{-3}\,s^{-1}$ and $3 \times 10^{+3}\,s^{-1}$, and for four different grain diameters: $d = 9.5,\,25,\,117$ and $315~m\mu$.  Their experimental results are shown by the points in Figs. \ref{HPFig1} and \ref{HPFig2} along with my theoretical fits to this data. As in previous work, the basic system parameters used in these equations are $T_P = 40 800$ K, $T_0 = 298$ K, $\mu_T = 1600$ MPa, and $\mu = 39.6$ GPa.

\begin{figure}[here]
\centering \epsfig{width=.5\textwidth,file=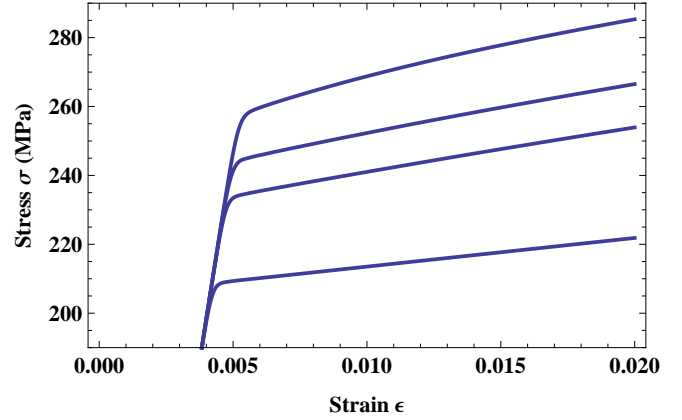} \caption{Enlarged graphs of the elastic-to-plastic transitions shown in Fig.\ref{HPyield1}. } \label{HPyield2}
\end{figure}
   
\begin{figure}[here]
\centering \epsfig{width=.5\textwidth,file=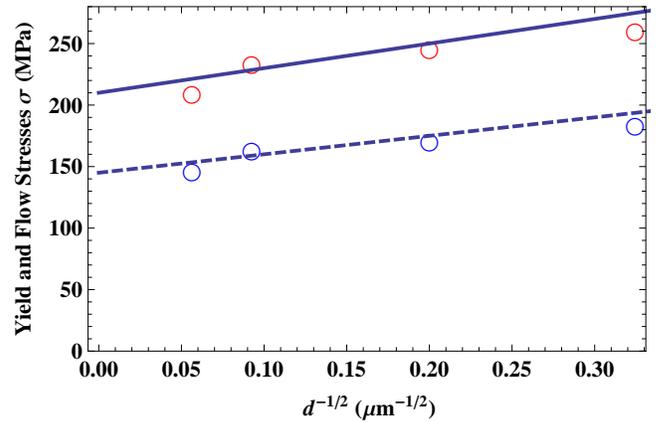} \caption{(Color online) Hall-Petch plots for the yield stresses shown in Figs. \ref{HPyield1} and  \ref{HPyield2} (upper solid curve), and for the flow stresses at $\epsilon = 0.1$ on the hardening curves shown in Fig.\ref{HPFig1} (lower dashed curve). } \label{HPyield3}
\end{figure}

I have slightly readjusted the parameters used in plotting the solid curves in Figs. \ref{HPFig1} and \ref{HPFig2} in order to focus on the fact that the conversion factor $\kappa_1$ in the equation of motion for $\tilde\rho$, Eq.(\ref{rhodot}), exhibits a strong and unambiguous Hall-Petch behavior.  Apparently, the stress concentrations at the edges of the grains, proportional to $d^{-1/2}$, amplify the strengths of the dislocation sources by factors as large as ten for the smallest grain sizes.  This effect is strain-rate dependent, so that the sequence of values of $\kappa_1$ is different for the high strain rate than it is for the low one.  These two HP-like behaviors are shown in Fig. \ref{HPkappa}. The analytic approximations shown by the solid and dashed curves are
\begin{equation}
\label{HPkappa1}
\kappa_1^{slow} \cong 2 + {21\over \sqrt{d}};~~~~\kappa_1^{fast} \cong 2 + {60\over \sqrt{d}}
\end{equation}
for the slow and fast cases respectively. Note that the rate dependence disappears at large grain sizes, in accord with the discussion in the paragraphs following Eq.(2.2) in \cite{JSL-16}.

This revised interpretation of the grain-size effects allows some simplification in evaluating other parameters.  I now find that the prefactor $\kappa_2$ in the equation of motion for the effective temperature $\tilde\chi$, Eq.(\ref{chidot}), is strain-rate dependent but approximately independent of the grain size, so that $\kappa_2 = 17$ for the slow case and $12$ for the fast one.  The steady-state values of the effective temperature $\tilde\chi_0$ seem to increase slightly with decreasing grain size, going from $0.240$ to $0.254$ for the small strain rate, and from $0.237$ to $0.250$ for the large one, consistent with the idea that the system becomes more disordered as the grains become smaller.  In all cases, I have chosen the initial value of $\tilde\rho$ to be $10^{-5}$.  Finally, I have chosen initial values of $\tilde\chi$ in the range $0.16\,-\,0.17$ in order to improve agreement with the early onset parts of these curves.  These last adjustments could indicate some variability in sample preparation.  These, and the other adjustments just mentioned, are well within my uncertainties in transcribing the published experimental data.  

\section{Simulated Loading Histories}

With the parameters determined here, I now can use the equations of motion for the state variables $\tilde\rho,\,\tilde\chi,\,\tilde\theta$ and the stress $\sigma$ to simulate a set of loading histories, and thereby look for Hall-Petch-like behaviors.  I do this, as in \cite{JSL-16a}, first by straining the samples slowly (at $\dot\epsilon = 10^{-3}\,s^{-1}$) up to $\epsilon = 0.1$, unloading them, and then reloading each of these prehardened samples rapidly (at $\dot\epsilon = 3 \times 10^{3}\,s^{-1}$).  In doing this theoretically, I use the final values of $\tilde\rho$ and $\tilde\chi$ from the first slow deformations as the initial values of those variables in the fast deformations.  

The first set of these experiments is plausibly realistic for Cu in the sense that I have turned off all of the ordinary thermal effects by setting $K = K_1 = 0$ in Eq.(\ref{thetadot}), so that the system remains at room temperature throughout the deformation, and thermal softening does not occur. (Equivalently, I could simply have chosen large values for $K_1$ and $K_2$.)  The results are shown in Fig.~\ref{HPyield1}.  Here we see four different yielding transitions corresponding to the four different grain sizes.  Enlarged graphs of these  transitions are shown in Fig.~\ref{HPyield2}, where we see more clearly that these are rapid, but smooth, transitions from elastic to plastic behavior.  The approximate values of the stress at these transitions are plotted as functions of $d^{-1/2}$ in Fig.~\ref{HPyield3}.  The Hall-Petch behavior is apparent here; but it is less pronounced than it is for $\kappa_1$ in Fig.~\ref{HPkappa}. The constant $\sigma_0$ part is much bigger than the stress-enhanced part.  As a consistency check, the dashed curve in Fig.~\ref{HPyield3} shows the flow stresses at $\epsilon = 0.1$ on the slow hardening curves shown in Fig.~\ref{HPFig1} as a function of $d$.  The resulting value of $k_s$ defined in Eq.(\ref{HPeqn}) is consistent with the corresponding point in Fig.~7a of \cite{MEYERSetal-95}.

\begin{figure}[here]
\centering \epsfig{width=.5\textwidth,file=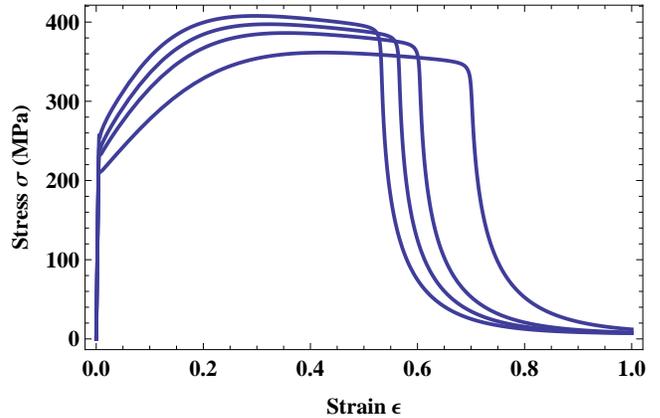} \caption{Stress-strain curves analogous to those shown in Fig.\ref{HPyield1}, but with thermal softening and an initial perturbation added to induce shear-banding failure.  Going from left to right, these failures occur for grain diameters $d = 9.5,\,25,\,117$ and $315~m\mu$.} \label{HP-ASB}
\end{figure}

\begin{figure}[here]
\centering \epsfig{width=.5\textwidth,file=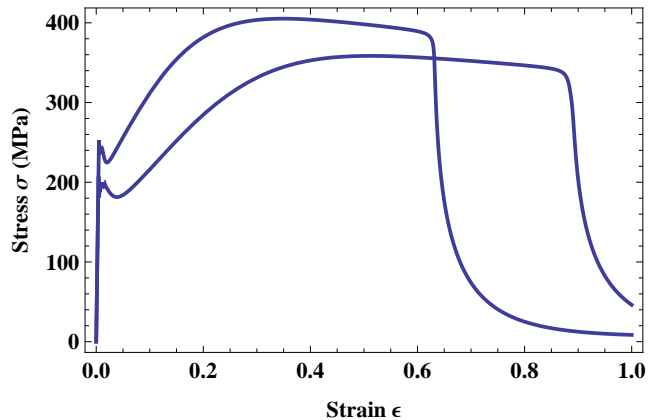} \caption{Stress-strain curves analogous to those shown in Fig.\ref{HP-ASB} for grain diameters $d = 9.5$ and $315~m\mu$, but with initial values of $\tilde\chi$ slightly reduced to simulate mild annealing between the slow prehardening and rapid reloading stages.  } \label{HP-ASBx}
\end{figure}

In a second set of theoretical experiments, I return to the ``pseudo copper'' that I introduced in \cite{JSL-16a}, but keep the grain-size dependent copper data used throughout this paper. That is, I now introduce in Eq.(\ref{thetadot}) a nonzero thermal conversion coefficient $K = 10^{-5}$ and a transport coefficient $K_2 = 10^{-9}$. For simplicity, I choose the lateral diffusion coefficient $K_1 = 0$, which produces a maximally sharp banding instability.  $K$ and $K_2$ are chosen solely so that that their effects are negligible at the small strain rate but appreciable at the larger one.  Also, as in  \cite{JSL-16a}, I introduce what I called a ``pseudo notch'' -- more accurately, a small perturbation along the $x$-axis of the sheared strip -- by writing
\begin{equation}
\tilde\chi(0, y) = \tilde\chi_i - \delta\,e^{- y^2/2\,y_0^2},
\end{equation}
where the $\tilde\chi_i$ are the same initial values of $\tilde\chi$ used previously and $\delta = 0.02$, $y_0 = 0.05$.  It is this weak, localized perturbation that triggers shear-banding instabilities.

The results of these experiments are shown in Fig.~\ref{HP-ASB}.  The initial yielding transitions are the same as those shown in Fig.~\ref{HPyield1}.  Here, however, these stress-strain curves show moderate thermal softening at intermediate strains and fail suddenly via shear-banding instabilities at large strains. The graphs of local strain rate and temperature near these shear bands look essentially the same as those shown in \cite{JSL-16a}. Note that the failure strains are larger for larger grain sizes and smaller for smaller ones.  In other words, the samples with larger grains are softer and more ductile; those with smaller grains are harder and more brittle. 

Finally, as a last theoretical experiment, I show in Fig.~\ref{HP-ASBx} what happens if, between the slow prehardening and fast reloading stages, I slightly decrease $\tilde\chi$ from its prehardened value.  That is, I simulate an intermediate  annealing step in which I slightly decrease the effective disorder temperature.  The resulting stress anomalies near $\epsilon = 0$ look much like those sometimes seen experimentally.  For example, this figure looks very much like Figure 8 in Marchand and Duffy's 1988 study of adiabatic shear banding in steel.  \cite{MARCHAND-DUFFY-88}     

\section{Remarks and Questions}

The most important assertion of this paper is that Hall-Petch effects arise, not primarily from resistance to dislocation flow at grain boundaries, but from enhanced creation of new dislocations at those places.  The more familar HP effects, such as increasing yield stresses, can be understood as indirect effects of the increased dislocation densities. 

There remain many unanswered questions.  For example, in \cite{JSL-15}, I showed that the rate-hardening anomaly reported in 1988 by Follansbee and Kocks \cite{FOLLANSBEE-KOCKS-88} can be understood simply by adding a linear strain-rate dependence to the coefficient $\kappa_1$; and I suggested that this rate dependence might be a grain-size effect.  How could this conjecture be tested and generalized? How should we write $\kappa_1$ as a function of both strain rate and grain size?  Similarly, what physical mechanism might explain why $\kappa_2$ in Eq.(\ref{chidot}) decreases as a function of strain rate?  Might the storage factor $\kappa_3$ in \cite{JSL-16a} play some role here? 

The direct comparisons with experiments in all of these papers (Refs. \cite{LBL-10,JSL-15,JSL-16,JSL-16a} and this one) are only for polycrystalline copper with grain sizes in the range of about $10 - 300$ microns.  This is far too narrow a basis for what I propose to be a general theory of polycrystalline plasticity.  Copper seems to be special in the sense that $\kappa_1$ can be measured directly, in effect by using Eq.~(\ref{Theta0}).  The important onset rate that appears in that equation is discussed in detail in \cite {KOCKS-MECKING-03}. Why is there no comparable information for other metals and alloys?  What differences in interpretation can we expect for polycrystalline solids with different crystalline symmetries?  I have not even mentioned, so far, the fact that the HP coefficient $k_s$ in Eq.(\ref{HPeqn}) may change sign when grain sizes become small of the order of nanometers. \cite{MEYERSetal-06} Why might this happen?  In short, I think that the ideas discussed here provide a new point of view from which to look at these questions, but only the beginnings of some answers.

These issues bring me back to some of the fundamental questions that I have been asking since the beginning  of this project.  Most importantly -- in looking at the large range of phenomena that seem to be relevant to  polycrystalline plasticity, how can we distinguish between causes and effects?  How can we determine whether an observed structural change such as the appearance of stacking faults or DRX grains is the cause of a qualitative change in behavior or simply a side effect of something else that is happening?  A more theoretically sophisticated version of this question is: What are the dynamically relevant state variables? 

Consider the strong assumption that was implicit in the way I simulated the loading histories that produced the yielding and failure curves in Figs.~\ref{HPyield1}, \ref{HP-ASB}, and \ref{HP-ASBx}.  I assumed that all of the memory of the pre-hardening deformations was carried by just two internal state variables, the density of dislocations $\tilde\rho$ and the effective disorder temperature $\tilde\chi$.  There are many other dynamical quantities that I could have -- and perhaps should have -- included.  The densities of stacking faults or DRX grains are good examples that appear often in the literature. Another would be some measure of the scale and intensity of cellular dislocation structures. The distributions of grain sizes might make important dynamical differences, especially when these sizes become small and grains begin to rearrange during deformation.  All of these quantities could be described by their own internal variables with their own equations of motion, and those variables and equations could be included in simulations of loading histories.  The only way to determine whether such extra ingredients are necessary is by careful experimentation.  

\begin{acknowledgments}

I would like to thank Professor Ronald Armstrong who, despite the fact that my results have contradicted many of his long-held opinions about dislocation physics, has continued to encourage me to continue my investigations.  He has been extremely helpful.  This research  was supported in part by the U.S. Department of Energy, Office of Basic Energy Sciences, Materials Science and Engineering Division, DE-AC05-00OR-22725, through a subcontract from Oak Ridge National Laboratory.

\end{acknowledgments}

\end{document}